\documentclass[a4paper]{jpconf}
\usepackage{graphicx}
\usepackage{amsmath}
\usepackage{amssymb}
\begin{document}
\title{A Time Domain Waveform for Testing General Relativity}

\author{C\'{e}dric Huwyler$^1$, Edward K. Porter$^2$, Philippe Jetzer$^1$}

\address{$^1$ Physik-Institut, Universit\"{a}t Z\"{u}rich, Winterthurerstrasse 190, 8057 Z\"{u}rich, Switzerland}
\address{$^2$ Fran\c{c}ois Arago Centre, APC, Universit\'e Paris Diderot, CNRS/IN2P3, CEA/Irfu, \\Observatoire de Paris, Sorbonne Paris Cit\'e, 10 rue A. Domon et L. Duquet, \\75205 Paris Cedex 13, France}

\ead{chuwyler@physik.uzh.ch}

\begin{abstract}
Gravitational-wave parameter estimation is only as good as the theory the waveform generation models are based upon.
It is therefore crucial to test General Relativity (GR) once data becomes available.
Many previous works, such as studies connected with the ppE framework by Yunes and Pretorius, rely on the stationary phase approximation (SPA)
to model deviations from GR in the frequency domain.
As Fast Fourier Transform algorithms have become considerably faster and in order to circumvent possible problems with the SPA,
we test GR with corrected time domain waveforms instead of SPA waveforms. Since a considerable amount of work has been done already in the field using SPA waveforms,
we establish a connection between leading-order-corrected waveforms in time and frequency domain, concentrating on phase-only  corrected terms.
In a Markov Chain Monte Carlo study, whose results are preliminary and will only be available later, we will assess the ability of the eLISA detector
to measure deviations from GR for signals coming from supermassive black hole inspirals using these corrected waveforms.
\end{abstract}

\section{Introduction}

This decade will witness the advent of direct gravitational wave (GW) detection: advanced ground-based detectors such as LIGO and Virgo are scheduled to come online as early
as 2015 and should hopefully provide numerous detections of stellar mass systems. Within the same time frame, pulsar timing array analysis is expected to advance
to a point where the first detection of supermassive black hole binaries (SMBHBs) in the nHz range will be possible.
By 2034, the European Space Agency plans to launch a space-based LISA-like GW detector called eLISA; the theme of
\textquotedblleft The Gravitational Universe\textquotedblright$\,$ was recently chosen for the ESA Cosmic Vision L3 mission in order to nourish the development of a space-based GW
mission.

The vast majority of algorithms in GW astronomy are based on the concept of matched filtering.  The detection of signals and the process of parameter estimation are conducted
 by assuming a particular underlying gravitational theory that is, a priori, considered
to be true; in our case this is the theory of general relativity (GR). That a waveform based on this theory is able to detect the signals present in the data, however, does
 not mean that it is the theory providing the best fit to the data. Alternative theories of gravity, deviating only slightly from GR, could also provide waveforms that also detect the signal just as well, albeit with different best-fit model parameters.
Thus, if we use the wrong waveform for parameter estimation, the inferred parameters will have a fundamental bias~\cite{yunespretorius2009} and will not fully reflect reality. 
This raises the question of
which particular theory of gravity should be preferred in the light of the measured data. By observing SMBHBs, which are among the strongest sources
of GWs, alternative theories can be compared to GR via a  Bayesian inference, where the odds ratio provides
a useful tool in measuring the betting odds for a certain alternative theory against GR. The space of available model hypotheses can then be populated with
\emph{specific} alternative theories (that are usually based on a modified Einstein-Hilbert action, such as scalar-tensor theories, Chern-Simons theory, etc.) or 
\emph{generic} features that are not present in GR (such as a non-zero graviton mass, Lorentz violation, variable $G(t)$, etc.). One could then test all these (classes of) theories
against GR. It is noteworthy that such approaches all have the common feature that they modify either the phase or the amplitude of the GW modelled by 
GR~\cite{yunespretorius2009}.
Interested in the mere fact of whether there exists a theory that describes the data better than GR, instead of testing for specific theories or generic features, one could thus 
just impose a set of theories that modify certain post-Newtonian (PN) amplitude or phase
coefficients of GR, without considering a particular action or a particular physical feature. One would then of course have to make sure that deviations originate
from fundamental physics and not from other effects, such as astrophysical pollution or pathologies in the chosen PN approximation scheme.

A considerable amount of work has been done in the aforementioned field. Arun, Iyer, Qusailah and Sathyprakash~\cite{arun2006} turned the question of testing GR
into the exact measurement of the PN coefficients $\Psi_i$ of the stationary phase
approximation (SPA) instead of measuring individual binary parameters, while Yunes and Pretorius~\cite{yunespretorius2009} added an amplitude correction to introduce their parameterized post-Einsteinian (ppE) scheme as 

\begin{equation}
 \label{eq:ppE}
 \tilde{h}(f) = \tilde{h}_\text{GR}(f) \, (1+\alpha \, u^a) \, e^{i\beta \, u^b},
\end{equation}
where tilde denotes a Fourier transform, $u = G\mathcal{M} \pi f/c^3$ is the reduced frequency, $G$ is the gravitational constant, $c$ is the speed of light
and $\mathcal{M} = M \eta^{3/5}$ is the chirp mass with total mass $M = m_1 + m_2$ and symmetric mass ratio $\eta = m_1 m_2 / M^2$. In this scheme, the ppE parameters
$(a,\alpha,b,\beta)$ can in general be real numbers. There are several studies regarding the performance of the ppE scheme in model selection 
\cite{cornishsampson2011,sampsonyunescornish2013,sampsoncornishyunes2014}. Sampson et al. have shown in \cite{sampsonyunescornish2013} that leading order corrections
as applied to GR in Eq.\ \eqref{eq:ppE} are already enough to distinguish GR from any competing theory. There is also the TIGER pipeline devised 
for ground-based detector data analysis \cite{lieatal2012}.

The previously mentioned approaches are constructed directly in the frequency domain and use the SPA to motivate the corrections they introduce to GR. The SPA was initially introduced into the field of 
GW astronomy as it saved on computational cost by making a Fast Fourier Transform (FFT) unnecessary, without much loss in accuracy. We chose not to rely on
the SPA for the following reasons: i) FFTs have nowadays become considerably faster and use only 3-8\% of the time needed for waveform generation, ii) several
studies have shown issues with the SPA, especially in the high-mass regime; there modifications such as the introduction of unphysical mass ratios or pseudo-4PN terms
are of some help \cite{pan2008, damour2000}, iii) even if the time-domain PN approximation breaks down by producing non-monotonic phase and frequency evolutions below a certain 
orbital separation, the
SPA waveforms still look fine at this point and do not indicate their apparent failure, and iv) as direct time-domain waveforms have been more commonly used to assess the parameter estimation capabilities of LISA-like
detectors.

In the following sections we will quickly review the waveform models in the time domain and the SPA, create a modified time domain waveform and compare it to a phase-only ppE scheme.

\section{\label{Sec:WaveformModels}Waveform Models}

\subsection{\label{Sec:TimeDomain}Time Domain}

In the time domain, the gravitational waveform is governed by the evolution of orbital frequency and phase \cite{blanchet2014}:
\begin{eqnarray}
  \label{eq:OmegaGR} \omega(\Theta) & = & \frac{c^3}{8 G M} \left[ \Theta^{-3/8} + \left(  \frac{743}{2688} + \frac{11}{32} \eta \right) \Theta^{-5/8} - \frac{3\pi}{10} \Theta^{-6/8}  \right. \nonumber \\
                                    &   & \left. +  \left(\frac{1855099}{14450688} + \frac{56975}{258048} \eta + \frac{371}{2048} \eta^2 \right) \Theta^{-7/8} \right], \\
  \label{eq:PhiGR} \Phi(\Theta) &=& \Phi_C - \frac{1}{\eta} \left[ \Theta^{5/8} + \left(  \frac{3715}{8064} + \frac{55}{96} \eta \right) \Theta^{3/8} - \frac{3\pi}{4} \Theta^{2/8} \right. \nonumber \\
                                    &   & \left. + \left( \frac{9275495}{14450688} + \frac{284875}{258048} \eta + \frac{1855}{2048} \eta^2 \right) \Theta^{1/8} \right], \\
  \nonumber
\end{eqnarray}
respectively. Here, $\Theta(t) = \frac{\eta c^3}{5 G M} (t_c-t)$ is the dimensionless time parameter, where $t_c$ is the formal time at coalescence.
The GW polarisations can then be written as
\begin{equation}
 \label{eq:hpluscross} 
 h_{+,\times} = \frac{2 G M \eta}{c^2 D_L} x\left[ H^{(0)}_{+,\times} + x^{1/2} H^{(1/2)}_{+,\times} + x H^{(1)}_{+,\times} + x^{3/2} H^{(3/2)}_{+,\times} + x^2 H^{(2)}_{+,\times} \right],
\end{equation}
where the post-Newtonian parameter $x = (GM\omega/c^3)^{2/3}$ is a function of the orbital frequency $\omega$, $D_L$ is the luminosity distance parameter and the $H^{(n)}_{+,\times}$
denote the harmonics of the phase which are proportional to sines and cosines of integer multiples of $\Phi$ and can be found in \cite{blanchet2014}. 
In the low-frequency approximation, which is the case for the vast majority of eLISA sources, the individual polarisations can be projected on the detector arm by using the antenna pattern
functions $F^{+,\times}_k$: $h_k(t) = F^+_k(t)\, h_+(t) + F^\times(t)\,  h_\times(t)$ for the $k$th channel of the detector.

\subsection{\label{Sec:FrequencyDomain}Frequency Domain}

Taking only the dominant harmonic into account, the Fourier transform of the time domain waveform is in the SPA expressed as
\begin{equation}
 \tilde{h}_k(f) = \mathcal{A}(f) \, e^{i\Psi(f)},
\end{equation}
where $\mathcal{A}(f) \propto f^{-7/6}$ and the SPA phase is defined to be
\begin{equation}
 \label{eq:PsiSPA}
 \Psi(f) = 2\pi f t(f) - 2\Phi[t(f)] - \frac{\pi}{4},
\end{equation}
or in terms of a 2PN expansion,
\begin{align}
 \Psi(f) &=  2\frac{t_c c^3}{G M} x^{3/2} - 2 \Phi_c - \frac{\pi}{4}  + \frac{3x^{-5/2}}{128\eta} \left[ 1 + \left( \frac{3715}{756} + \frac{55\eta}{9}\right) x - 16\pi x^{3/2} \right. \nonumber \\
         & \left. +\left( \frac{15293365}{508032} + \frac{27145\eta}{504} + \frac{3085\eta^2}{72} \right) x^2 \right].\nonumber \\
\end{align}

\section{Modified Waveforms}

In contrast to \cite{yunespretorius2009}, we chose to take only modifications to the phase into account and neglect any corrections to the amplitude (i.e. $a=\alpha=0$).
Eq. \eqref{eq:ppE} amounts to a modification of the SPA phase in terms of
\begin{equation}
 \label{eq:PsiPPE}
 \Psi_\text{NGR}(b,\beta; \, u) = \Psi_\text{GR}(u) + \beta \, u^b.
\end{equation}
where `NGR' stands for `non-GR'. In a similar way, however slightly more adapted to Eq.\ \eqref{eq:ppE}, we chose to modify the time domain phase
\begin{equation}
 \label{eq:PhiModAlpha}
 \Phi_\text{NGR}(i,\kappa; \, \Theta) = \Phi_\text{GR}(\Theta) - \frac{1}{\eta} \kappa \, \Theta^{\frac{5-2i}{8}},
\end{equation}
with $\kappa \in \mathbb{R}$. If leading order corrections are enough to detect a deviation from GR in the frequency domain in Eq.\ \eqref{eq:PsiPPE}, then in principle
they should also be sufficient in Eq.\ \eqref{eq:PhiModAlpha}. For the sake of a comparison between corrected time domain and SPA waveforms, however, we include higher orders to 
Eq.\ \eqref{eq:PhiModAlpha} such that
\begin{equation}
\label{eq:modelPhi}
 \Phi_{\text{NGR}}(i,\kappa; \, \Theta) =  \Phi_{\text{GR}}(\Theta) - \frac{1}{\eta} \sum_i \kappa_i \Theta^{\frac{5-2i}{8}}.
\end{equation}
Our aim is now to fix the coefficients $\kappa_i$ in such a way that the Fourier transform of a waveform with an orbital phase given by Eq.\ \eqref{eq:modelPhi} is as close as possible 
to an SPA waveform with a phase described by Eq.\ \eqref{eq:PsiPPE}. To this end, let us construct a relation which allows us to compare the coefficients of Eqs.\ \eqref{eq:PsiPPE} and \eqref{eq:modelPhi}
by taking the derivative of the definition of the SPA phase. Noting that $\frac{d\Phi}{du}(u) = \frac{dt}{du} \frac{c^3}{G\mathcal{M}} u$, we arrive
at
\begin{table}[b]
\begin{center}
\caption{\label{Table:kappabeta} An overview of how the corrections $\kappa_i(b,\beta)$ have to be fixed in order to approximately represent a leading order phase-only ppE waveform after a Fourier transform.}
\lineup
\begin{tabular}{l|lllll}
\br
b	  & -5/3 & -4/3 & -1 & -2/3 & -1/3 \\
\hline
$\kappa_{0}$  			& $16 \beta$ & 0 & 0 & 0 & 0 \\
%\hline
$\kappa_{1/2}$  		& 0 & $8 \beta \eta^{1/5}$ & 0 & 0 & 0  \\
%\hline
$\kappa_{1}$  			& $-16 \beta \Phi_1$ & 0 & $4 \beta \eta^{2/5}$ & 0 & 0 \\
%\hline
$\kappa_{3/2}$  		& $-\frac{32}{3} \beta \Phi_{3/2}$ &  $-\frac{32}{5} \beta \Phi_1 \eta^{1/5}$ & 0 & $2\beta \eta^{3/5}$ & 0 \\
%\hline
$\kappa_{2}$ 	 		& $16 \beta \left(\frac{4}{5}\Phi_1^2 - \frac{1}{3}\Phi_2 \right)$ & $-\frac{64}{15} \beta \Phi_{3/2} \eta^{1/5}$ & $-\frac{12}{5} \beta \Phi_1 \eta^{2/5}$ & 0 & $\beta \eta^{4/5}$ \\
\br
\end{tabular}
\end{center}
\end{table}

\begin{equation}
 \frac{d\Psi}{du} = 2 \frac{c^3}{G \mathcal{M}}\,  t(u).
\end{equation}
This enables us to write the time-of-frequency and the frequency-of-time functions in the simple form
\begin{equation}
\label{tu}
 t(u) = \frac{1}{2} \frac{G\mathcal{M}}{c^3} \frac{d\Psi}{du}, \qquad  u(t) = \frac{G\mathcal{M}}{c^3} \frac{d\Phi}{dt}.
\end{equation}
Since one requires $u[t(u)] = u$, the coefficients of $\Psi(u)$ given the coefficients of $\Phi(\Theta)$ can be computed by evaluating
\begin{equation}
 \label{eq:utu}
 u[ \Theta( u ) ]_{\text{2PN}} = u \, \left(1+\sum_{k=0}^4 u^{k/3} \mathcal{A}_k \right) = u,
\end{equation}
expanded up to 2PN order in $u$. Assuming that $\kappa_i$ and $\beta$ are small enough such that all terms can be expanded at linear order, and by setting
$\mathcal{A}_k$ to zero results in a linear system that can be solved for $\kappa_i(b,\beta)$. We chose to fix the correction orders between 0PN and 2PN, in particular,
 $b \in \{-5/3, -4/3, -1, -2/3, -1/3\}$ and $i \in \{0,1/2,1,3/2,2\}$; this corresponds to 0PN, 0.5PN, 1PN, 1.5PN and 2PN corrections to the phase/frequency, respectively.
For simplicity, we chose not to consider `negative' PN terms due to dipole radiation because we plan to apply the waveforms to SMBH inspirals where, at least in scalar-tensor theories,
there appears to be no dipole radiation present \cite{lang2014}.
Subsequently, we fix one particular value of $b$ at a time and allow the full sum of corrections proportional to $\kappa_i$ in the 
time domain phase. We have solved the linear system and have computed $\kappa_i(b,\beta)$ for each value of $b$. The results are listed in table \ref{Table:kappabeta}.
In the table, some patterns are clearly visible:
Not surprisingly, the $\kappa_i$ are always proportional to $\beta$.
For a frequency domain phase correction entering at $n$PN order, $n=(3b+5)/2$, $\kappa_{i<n} = 0$ and the $\kappa_i$ 
are proportional to $\eta^{\frac{2n}{5}}$. Furthermore, the lowest order correction $\kappa_i$ has the numeric pre-factor 
$2^{4-2i}$.
In the first off-diagonal element, the coefficients are proportional to $\Phi_0=1$, in the secondary off-diagonal to $\Phi_{1/2} = 0$, 
in the third to $\Phi_1$ and in the $4^{\text{th}}$ to $\Phi_{3/2}$. In the  $5^{\text{th}}$ off-diagonal, terms proportional
both to $\Phi_1^2$ and $\Phi_2$ can appear. 

As previously mentioned, we chose to consider only the leading order corrections and are thus able to write the corrected time domain phase as
(using $i=(3b+5)/2$)
\begin{equation}
 \label{eq:PhiModBeta}
 \Phi_{\text{NGR}}^{(\pm)}(b,\beta;\,\Theta) = \Phi_\text{GR}(\Theta) \pm  2^{-1-3b} \beta \, \eta^{3b/5} \Theta^{-3b/8},
\end{equation}
with a modified orbital frequency given by
\begin{equation}
 \label{eq:OmegaModBeta}
 \omega_{\text{NGR}}^{(\pm)}(b,\beta;\,\Theta) = \omega_\text{GR}(\Theta) \pm  2^{-4-3b} \frac{3\beta}{5} \frac{c^3}{GM}\, \eta^{3b/5+1} \Theta^{-3b/8-1}.
\end{equation}
Equations \eqref{eq:PhiModBeta} and \eqref{eq:OmegaModBeta} can then be used in the GW polarisations described by Eqs.\ \eqref{eq:hpluscross} to generate a modified waveform that approximately uses the same $\beta$ parameter as a leading order phase-only ppE scheme. We assume that our coupling constant $\beta$ is 
manifestly positive as we would like to distinguish between possible positive and negative corrections to the GR waveform.
\section{Summary}

We have introduced alternative theory corrections to the time domain phase and orbital frequency for a GW generated by a compact binary
inspiral in GR. In order to compare these waveforms to leading order phase-corrected frequency domain waveforms, we have established
a relation between the magnitude parameters $\kappa$ and $\beta$.
We have implemented the corrected time domain waveforms into a Markov Chain Monte Carlo code that assesses the ability of the eLISA
detector to discriminate between GR and alternatives theories in the context of SMBHB inspirals. Preliminary results are available, but will be published in \cite{huwyler2014}.

\end{document}